\documentclass[10pt,conference]{IEEEtran}
\IEEEoverridecommandlockouts

\usepackage{cite}
\usepackage{amsmath,amssymb,amsfonts}
\usepackage{algorithmic}
\usepackage{graphicx}
\usepackage{textcomp}
\usepackage{xcolor}
\usepackage{balance}
\usepackage{subfigure} 
\usepackage{amsmath}
\usepackage{subcaption}
\usepackage{algorithm}
\usepackage{caption}
\usepackage{geometry}
\usepackage{algorithm}
\usepackage{subcaption}
\usepackage{amsthm}

\newtheorem{Remark}{Remark}

\usepackage{stfloats}

\begin{document}
\captionsetup[figure]{labelformat={default},labelsep=period}
\title{Joint Beamforming for Backscatter Integrated Sensing and Communication}
\author{Zongyao Zhao$^{1,2}$, Tiankuo Wei$^{1,2}$, Zhenyu Liu$^{1}$, Xinke Tang$^{2}$, Xiao-Ping Zhang$^{1}$, Yuhan Dong$^{1,2,*}$\\
	$^1$Shenzhen International Graduate School, Tsinghua University, Shenzhen, P. R. China\\
    $^2$Pengcheng Laboratory, Shenzhen, P. R. China\\
	Email: zhaozong21@mails.tsinghua.edu.cn, 
              wtk23@mails.tsinghua.edu.cn,
              zhenyuliu@sz.tsinghua.edu.cn,\\
              tangxk@pcl.ac.cn, 
               xiaoping.zhang@sz.tsinghua.edu.cn,
               dongyuhan@sz.tsinghua.edu.cn}

 \maketitle

\begin{abstract}
Integrated sensing and communication (ISAC) is a key technology of next generation wireless communication. Backscatter communication (BackCom) plays an important role for internet of things (IoT). Then the integration of ISAC with BackCom technology enables low-power data transmission while enhancing the system sensing ability, which is expected to provide a potentially revolutionary solution for IoT applications. In this paper, we propose a novel backscatter-ISAC (B-ISAC) system and focus on the joint beamforming design for the system. We formulate the communication and sensing model of the B-ISAC system and derive the metrics of communication and sensing performance respectively, i.e., communication rate and detection probability. We propose a joint beamforming scheme aiming to optimize the communication rate under sensing constraint and power budget. A successive convex approximation (SCA) based algorithm and an iterative algorithm are developed for solving the complicated non-convex optimization problem. Numerical results validate the effectiveness of the proposed scheme and associated algorithms.  The proposed B-ISAC system has broad application prospect in IoT scenarios.
\end{abstract}


\section{Introduction}
Integrated sensing and communication (ISAC) technology has recently emerged as a candidate technology of the next generation wireless network, which aims to integrate sensing and communication into one system to improve spectrum efficiency and hardware efficiency while providing sensing and communication service simultaneously\cite{Hassan2016,Liu2023,LiuX2020,Zhao2024,Zhao2022}. As a promising technology for low-power communication, backscatter communication (BackCom) uses radio frequency (RF) tags to enable passive communication links by scattering RF signals to the reading device \cite{Niu2019,Jiang2023}. 

The combination of BackCom with ISAC promises to combine the advantages of both technologies to enable energy-efficient passive communication while improving the sensing performance, especially for internet of things (IoT) applications. The authors in \cite{Luo2023} further considered the detection requirement of RF tags and developed a joint beamforming design for ISAC system with backscatter tags, which minimize the total transmit power while meeting the tag detection and communication requirements. In summary, research on on joint beamforming schemes for backscatter integrated sensing and communication is scarce and thus requires further exploration of this novel paradigm. Moreover, existing schemes only focus on the optimization of the energy consumption, however the optimization of communication rate is still remained to be explored.

In this paper, we propose a novel backscatter-ISAC \mbox{(B-ISAC)} system and design a joint beamforming scheme to optimize communication rate under the sensing constraint and power budget. Specifically, we model the signals received at the tag, user equipment (UE), and access point (AP). Further, we establish the communication and sensing model and derive the metrics of communication and sensing performance respectively, i.e., communication rate and detection probability. Then, we propose a joint beamforming scheme aiming to optimize the UE communication rate while ensuring the tag is working normally. We develop a successive convex approximation (SCA) based algorithm and an alternating algorithm for the joint beamforming scheme. Extensive simulation validate the effectiveness of the proposed scheme and associated algorithms. 

The proposed B-ISAC system fully utilizes the passive back scatter
properties of RF tags to achieve low-power communication and sensing capabilities. It is expected to has broad application
prospect in IoT scenarios.

The remainder of this paper is organized as follows. Sec.~\ref{sec2} introduces the B-ISAC system and signal model. The joint beamforming scheme for communication rate optimization is proposed in Sec.~\ref{sec3}. Numerical results are presented in Sec.~\ref{sec5}. Finally, the conclusions are drawn in Sec.~\ref{sec6}.

\emph{Notation}: In this paper, boldface lower-case and upper-case letters denote vectors and matrices respectively. $\mathbb{R}$ and $\mathbb{C}$ represent the real and complex sets respectively. $|\cdot|$, $||\cdot||$, and $||\cdot||_{F}$ are  absolute value, Euclidean norm, and Frobenius norm, respectively. $\left( \cdot \right)^{-1}$ and $\left( \cdot \right)^{\dagger}$ denote the inverse and pseudo inverse, respectively. $\left( \cdot \right)^T$, $\left( \cdot\right)^*$, and $\left( \cdot \right)^H$ represent transpose, complex conjugate, and Hermitian transpose, respectively. $\mathbb{E}\left( \cdot \right)$ represents statistical expectation. $\mathrm{Re}\left\{ \cdot\right\}$ returns the real part of a complex number. $j$ is the imaginary unit, which means $j^2=-1$. $\mathbf{I}_N$ is the $N\times N$ identity matrix. $\mathbf{1}=\left[1,1,\ldots,1\right]^T\in\mathbb{R}^N$. $\mathbf{A}\succeq0$ means that $\mathbf{A}$ is a positive semidefinite matrix. $\odot$ represents the Hadamard product. $\mathrm{diag}\left(\mathbf{a} \right)$ returns a diagonal matrix, the vector composed of its diagonal elements is $\mathbf{a}$. $\mathrm{Tr}\left(\mathbf{A} \right)$ and $\mathrm{rank}\left(\mathbf{A}\right)$ compute the trace and rank of matrix $\mathbf{A}$ respectively. $\mathrm{chol}\left(\mathbf{A}\right)$ returns the Cholesky decomposition of matrix $\mathbf{A}$. $\mathrm{vec}\left(\mathbf{A}\right)$ vectorizes matrix $\mathbf{A}$ by column-stacking.
\section{System and Signal Model} \label{sec2}
 \subsection{System Model}
 \begin{figure}[!t]
 \captionsetup{font=small}
\centering
\includegraphics[width=0.28\textwidth]{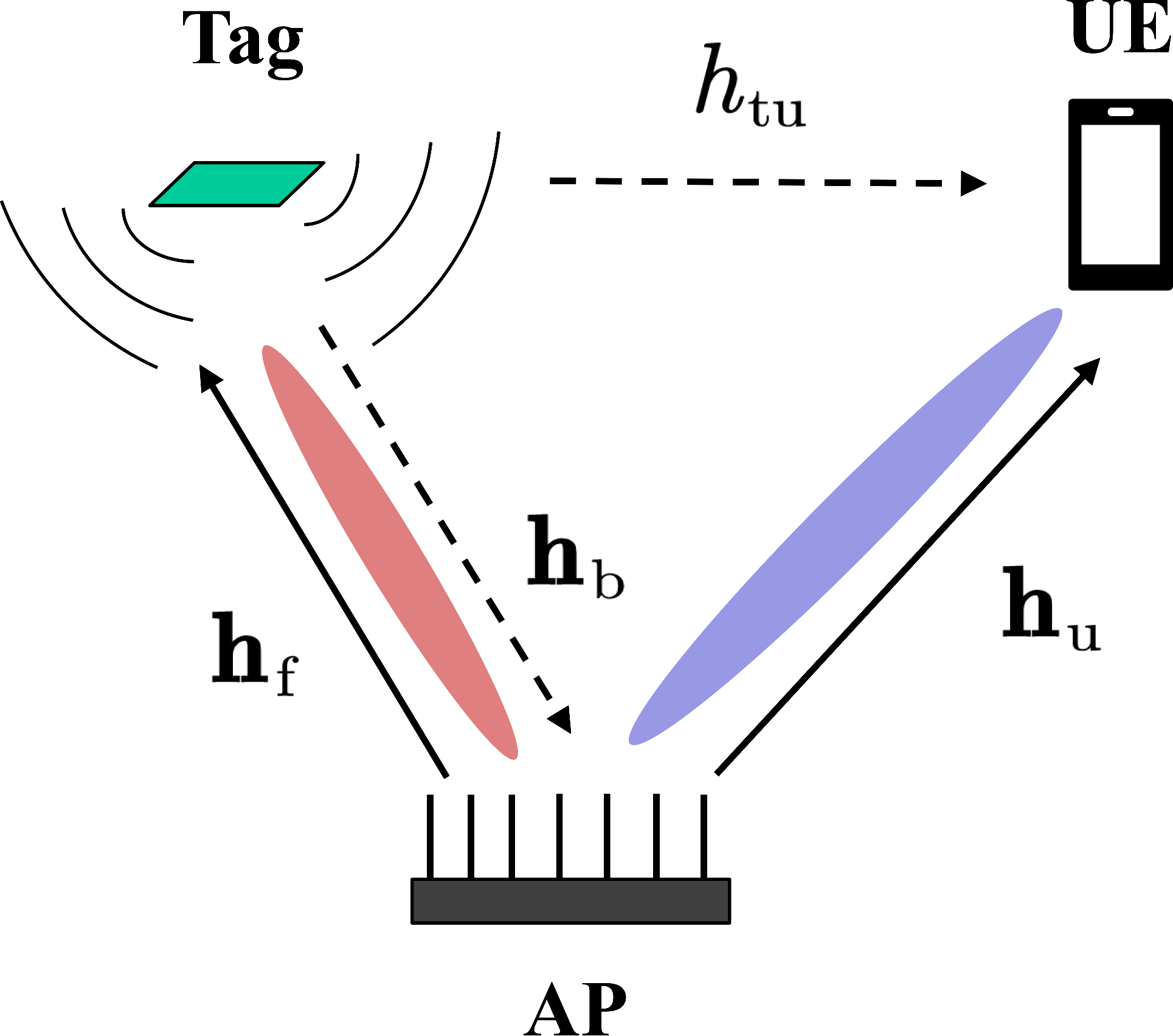}
\caption{B-ISAC System Model.}
\label{Fig1}
\vspace{-0.5cm} 
\end{figure}
As shown in Fig.~\ref{Fig1}, the B-ISAC system is composed of an AP, a UE, and a passive tag. The AP facilitates communication service to the UE, while providing sensing and communication support for the tag. The tag receives the signal transmitted by the AP to decode downlink data and modulates the uplink data onto the backscattered signal to achieve two-way data transmission. At the same time, the AP receives the backscattered signal and carries out further processes to obtain sensing information, such as position, status and etc. For the UE, the backscattered signal from the tag can be regarded as interference.  

It is assumed that the AP is equipped with uniform linear transmit and receive arrays with $N_t$ and $N_r$ antenna elements, respectively. There is $\lambda/2$ inter-spacing between neighboring antenna elements, where $\lambda$ is the carrier wavelength. Without loss of generality, we assume $N_t \leqslant N_r$. The tag and UE are equipped with a single antenna, respectively. It is also assumed that the self-interference between the transmit and receive arrays can be ignored.

$\mathbf{X}\in \mathbb{C} ^{{N_t}\times L}$ denote the transmitted signal at the AP, where $L$ is the length of the transmitted signal such that $L>N_t$. $\mathbf{X}$ can be expressed as
\begin{gather}
\begin{aligned}
\mathbf{X}=\mathbf{WS}=\mathbf{w}_{\mathrm{u}}\mathbf{s}_{\mathrm{u}}^{H}+\mathbf{w}_{\mathrm{t}}\mathbf{s}_{\mathrm{t}}^{H}+\mathbf{W}_{\mathrm{s}}\mathbf{S}_{\mathrm{s}}\in \mathbb{C} ^{N_t\times L},
\label{eq1}
\end{aligned}
\end{gather}
where the joint beamforming matrix $\mathbf{W}$ and the data augmentation matrix $\mathbf{S}$ are respectively given by
\begin{align}
\mathbf{W}&=\left[\mathbf{w}_{\mathrm{u}},\mathbf{w}_{\mathrm{t}},\mathbf{w}_{1},\mathbf{w}_{2},\ldots,\mathbf{w}_{N_t} \right] \in \mathbb{C} ^{N_t\times (N_t+2)},\\
\mathbf{S}&=\left[ \mathbf{s}_{\mathrm{u}},\mathbf{s}_{\mathrm{t}},\mathbf{s}_1, \mathbf{s}_2 \ldots,\mathbf{s}_{N_t} \right] ^H\in \mathbb{C} ^{{(N_t+2)}\times L},
\end{align}
where $\mathbf{w}_{\mathrm{u}}\in \mathbb{C} ^{N_t\times {1}}$ and $\mathbf{w}_{\mathrm{t}}\in \mathbb{C} ^{N_t\times {1}}$ are beamforming vectors for the UE and tag, respectively. Vectors $\mathbf{s}_{\mathrm{u}}\in \mathbb{C} ^{L\times {1}}$ and $\mathbf{s}_{\mathrm{t}}\in \mathbb{C} ^{L\times {1}}$ are data streams for the UE and tag, respectively. The additional probing stream $\left[\mathbf{s}_1,\mathbf{s}_2,...,\mathbf{s}_{N_t} \right]^H =\mathbf{S}_{\mathrm{s}}\in \mathbb{C}^{N_t\times L}$  is introduced to extend the sensing degrees of freedom (DoF) of the transmit waveform. In this way, the DoF of the transmit waveform can be extended to its maximum value \cite{LiuX2020}. Note that these dedicated probing streams are deterministic and do not contain any information. Moreover, $\left[\mathbf{w}_1,\mathbf{w}_2,...,\mathbf{w}_{N_t} \right] =\mathbf{W}_{\mathrm{s}}\in \mathbb{C}^{N_t\times N_t}$ is the dedicated auxiliary beamfoming matrix corresponding to the dedicated probing streams. 

It is assumed that there is no correlation between dedicated probing streams $\mathbf{S}_{\mathrm{s}}$ and data streams $[\mathbf{s}_{\mathrm{u}},\mathbf{s}_{\mathrm{t}}]^H$. When $L$ is sufficiently large, the random sample covariance matrix of the random communication data streams $[\mathbf{s}_{\mathrm{u}},\mathbf{s}_{\mathrm{t}}]^H$ can be approximately considered to satisfy $\frac{1}{L}[\mathbf{s}_{\mathrm{u}},\mathbf{s}_{\mathrm{t}}]^H[\mathbf{s}_{\mathrm{u}},\mathbf{s}_{\mathrm{t}}] \approx \mathbf{I}_{2}$. 
At the same time, the dedicated probing streams $\mathbf{S}_{\mathrm{s}}$ is also elaborated to satisfy $\frac{1}{L}\mathbf{S}_{\mathrm{s}}{\mathbf{S}_{\mathrm{s}}}^H=\mathbf{I}_{N_t}$. Therefore, $\mathbf{S}$ satisfies
\begin{gather}
\begin{aligned}
\frac{1}{L}\mathbf{SS}^H\approx\mathbf{I}_{N_t+2}.
\end{aligned}
\end{gather}
The sample covariance matrix of waveform $\mathbf{X}$ is given by
\begin{gather}
\begin{aligned}
\mathbf{R}_\mathbf{X}=\frac{1}{L}\mathbf{XX}^H\approx\mathbf{WW}^H\in \mathbb{C} ^{N_t \times N_t},
\label{eq5}
\end{aligned}
\end{gather}
Then, the beam pattern of the transmit waveform is
\begin{gather}
\begin{aligned}
P\left( \theta \right) = \mathbf{a}^H\left( \theta \right) \mathbf{R}_\mathbf{X}\mathbf{a}\left( \theta \right),
\label{eq6}
\end{aligned}
\end{gather}
where $\mathbf{a}\left( \theta \right)= \left[ 1,e^{j\pi \sin \theta},e^{2j\pi \sin \theta},...,e^{\left( N_t-1 \right) j\pi \sin \theta} \right] ^T \in \mathbb{C} ^{1 \times N_t}$ represents the steering vector of the transmit array, and $\theta$ represents the azimuth angle with respect to the array.
\subsection{Backscatter Model}
The signal $\mathbf{y}_{\mathrm{t}}\in \mathbb{C}^{1\times L}$ received by the tag is given by
\begin{align}
\mathbf{y}_{\mathrm{t}}&=\mathbf{h}_{\mathrm{f}}\mathbf{X}+\mathbf{n}_{\mathrm{t}} \nonumber
\\
&=\mathbf{h}_{\mathrm{f}}\mathbf{w}_{\mathrm{u}}\mathbf{s}_{\mathrm{u}}^{H}+\mathbf{h}_{\mathrm{f}}\mathbf{w}_{\mathrm{t}}\mathbf{s}_{\mathrm{t}}^{H}+\mathbf{h}_{\mathrm{f}}\mathbf{W}_{\mathbf{s}}\mathbf{S}_{\mathbf{s}}+\mathbf{n}_{\mathrm{t}},
\label{eq7}
\end{align}
where $\mathbf{h}_{\mathrm{f}}\in \mathbb{C} ^{1\times N_t}$ denotes the channel from the AP to tag, $\mathbf{n}_{\mathrm{t}}\in \mathbb{C} ^{1\times L}$ is the receiver noise vector at the tag, which is assumed to follow a zero-mean complex Gaussian distribution as $\mathbf{n}_{\mathrm{t}}\sim \mathcal C\mathcal N \left( 0,\sigma_{\mathrm{t}}^{2}\mathbf{I}_L \right)$. On the right hand side of (\ref{eq7}), the first part $\mathbf{h}_{\mathrm{f}}\mathbf{w}_{\mathrm{u}}\mathbf{s}_{\mathrm{u}}$ represents the interference caused by the UE communication data, the second part $\mathbf{h}_{\mathrm{f}}\mathbf{w}_{\mathrm{t}}\mathbf{s}_{\mathrm{t}}$ represents the desired signal for the tag, the third part $\mathbf{h}_{\mathrm{f}}\mathbf{W}_{\mathbf{s}}\mathbf{S}_{\mathbf{s}}$ represents the interference caused by the dedicated probing stream, and $\mathbf{n}_{\mathrm{t}}$ is the noise vector. Therefore, the signal-to-interference-plus-noise ratio (SINR) of the signal received at the tag can be expressed as
\begin{align}
\gamma _{\mathrm{t}}&=\frac{\mathbb{E} \left( \|\mathbf{h}_{\mathrm{f}}\mathbf{w}_{\mathrm{t}}\mathbf{s}_{\mathrm{t}}^{H}\|^2 \right)}{\mathbb{E} \left( \|\mathbf{h}_{\mathrm{f}}\mathbf{w}_{\mathrm{u}}\mathbf{s}_{\mathrm{u}}^{H}\|^2 \right) +\mathbb{E} \left( \left\| \mathbf{h}_{\mathrm{f}}\mathbf{W}_{\mathrm{s}}\mathbf{S}_{\mathrm{s}} \right\| ^2 \right) +\mathbb{E} \left( \|\mathbf{n}_{\mathrm{t}}\|^2 \right)} \nonumber
\\ 
&=\frac{|\mathbf{h}_{\mathrm{f}}\mathbf{w}_{\mathrm{t}}|^2}{|\mathbf{h}_{\mathrm{f}}\mathbf{w}_{\mathrm{u}}|^2+\left\| \mathbf{h}_{\mathrm{f}}\mathbf{W}_{\mathrm{s}} \right\| ^2+\sigma _{\mathrm{t}}^{2}}.
\label{eq8}
\end{align}
Then, the tag modulates the uplink data onto the backscattered signal. The remodulated backscattered signal $\mathbf{y}_{\mathrm{b}}\in \mathbb{C} ^{1\times L}$ reflected by the tag is given by
\begin{align}
\mathbf{y}_{\mathrm{b}}=&\sqrt{\alpha}\mathbf{y}_{\mathrm{t}} \odot  \mathbf{c}_{\mathrm{t}}\nonumber
 \\
=&\sqrt{\alpha}\mathbf{h}_{\mathrm{f}}\mathbf{X}\odot \mathbf{c}_{\mathrm{t}}+\sqrt{\alpha}\mathbf{n}_{\mathrm{t}}\odot  \mathbf{c}_{\mathrm{t}}
\end{align}
where $\alpha$ is the backscatter modulation efficiency coefficient, and $\mathbf{c}_t$ is the uplink data satisfying $\mathbb{E} [|\mathbf{c}_t|^2] =1 $.

Vector $\mathbf{h}_{\mathrm{b}}\in \mathbb{C} ^{N_r\times 1 }$ represents the channel from the tag to AP. Then, the backscattered signal received at the AP $\mathbf{Y}_{\mathrm{ap}} \in \mathbb{C} ^{N_r \times L} $ can be expressed as
\begin{align}
\mathbf{Y}_{\mathrm{ap}}=&\mathbf{h}_{\mathrm{b}}\mathbf{y}_{\mathrm{b}}+\mathbf{N}_{\mathrm{ap}}\nonumber
\\
=&\sqrt{\alpha}\mathbf{h}_{\mathrm{b}}\mathbf{h}_{\mathrm{f}}\mathbf{X}\odot \mathbf{c}_{\mathrm{t}}+\sqrt{\alpha}\mathbf{h}_{\mathrm{b}}\mathbf{n}_{\mathrm{t}}\odot \mathbf{c}_{\mathrm{t}}+\mathbf{N}_{\mathrm{ap}}
\end{align}
where $\mathbf{N}_{\mathrm{ap}}\in \mathbb{C} ^{N_r\times L}$ is the noise matrix at the AP, which is assumed as $\mathrm{vec}(\mathbf{N}_{\mathrm{ap}})\sim \mathcal C\mathcal N \left( 0,\sigma_{\mathrm{ap}}^{2}\mathbf{I}_{{N_r}L} \right)$. Then, we use a receive combine vector $\mathbf{w}_{\mathrm{r}}\in \mathbb{C} ^{1\times N_r}$ to combine the signal of $N_r$ antennas. The combined signal $\tilde{\mathbf{y}}_{\mathrm{ap}} \in \mathbb{C} ^{1\times L}$ is given by
\begin{align}
\tilde{\mathbf{y}}_{\mathrm{ap}}=&\mathbf{w}_{\mathrm{r}}\mathbf{Y}_{\mathrm{ap}}\nonumber
\\
=&\sqrt{\alpha}\mathbf{w}_{\mathrm{r}}\mathbf{h}_{\mathrm{b}}\mathbf{h}_{\mathrm{f}}\mathbf{X}\odot \mathbf{c}_{\mathrm{t}}+\sqrt{\alpha}\mathbf{w}_{\mathrm{r}}\mathbf{h}_{\mathrm{b}}\mathbf{n}_{\mathrm{t}}\odot \mathbf{c}_{\mathrm{t}}+\mathbf{w}_{\mathrm{r}}\mathbf{N}_{\mathrm{ap}}.
\label{eq11}
\end{align}
It is worth noting that the transmit signal $\mathbf{X}$ is completely known to the AP, so 
the uplink data can be decoded using $\mathbf{X}$. We denote $\widetilde{\mathbf{C}}_{\mathrm{t}}=\mathrm{diag}\left( \mathbf{c}_{\mathrm{t}} \right)$. Consequently, the received SINR at the AP is 
\begin{align}
\gamma _{\mathrm{ap}}&=\frac{\mathbb{E} \left( \left\| \sqrt{\alpha}\mathbf{w}_{\mathrm{r}}\mathbf{h}_{\mathrm{b}}\mathbf{h}_{\mathrm{f}}\mathbf{X}\odot \mathbf{c}_{\mathrm{t}} \right\| ^2 \right)}{\mathbb{E} \left( \left\| \sqrt{\alpha}\mathbf{w}_{\mathrm{r}}\mathbf{h}_{\mathrm{b}}\mathbf{n}_{\mathrm{t}}\odot \mathbf{c}_{\mathrm{t}} \right\| ^2 \right) +\mathbb{E} \left( \left\| \mathbf{w}_{\mathrm{r}}\mathbf{N}_{\mathrm{AP}} \right\| ^2 \right)}\nonumber
\\
&=\frac{\alpha \mathbb{E} \left( \mathbf{w}_{\mathrm{r}}\mathbf{h}_{\mathrm{b}}\mathbf{h}_{\mathrm{f}}\mathbf{X}\widetilde{\mathbf{C}}_{\mathrm{t}} \widetilde{\mathbf{C}}_{\mathrm{t}} ^H\mathbf{X}^H\mathbf{h}_{\mathrm{f}}^{H}\mathbf{h}_{\mathrm{b}}^{H}\mathbf{w}_{\mathrm{r}}^{H} \right)}{\alpha \mathbb{E} \left( |\mathbf{w}_{\mathrm{r}}\mathbf{h}_{\mathrm{b}}\mathbf{n}_{\mathrm{t}}\widetilde{\mathbf{C}}_{\mathrm{t}} \widetilde{\mathbf{C}}_{\mathrm{t}}^H\mathbf{n}_{\mathrm{t}}^{H}\mathbf{h}_{\mathrm{b}}^{H}\mathbf{w}_{\mathrm{r}}^{H}| \right) +\left\| \mathbf{w}_{\mathrm{r}} \right\| ^2\sigma _{\mathrm{ap}}^{2}}\nonumber
\\
&\overset{\left( a \right)}{\approx}\frac{\alpha  \mathbf{w}_{\mathrm{r}}\mathbf{h}_{\mathrm{b}}\mathbf{h}_{\mathrm{f}}\mathbf{W}\mathbf{W}^{H}\mathbf{h}_{\mathrm{f}}^{H}\mathbf{h}_{\mathrm{b}}^{H}\mathbf{w}_{\mathrm{r}}^{H}}{\alpha |\mathbf{w}_{\mathrm{r}}\mathbf{h}_{\mathrm{b}}|^2\sigma _{\mathrm{t}}^{2}+\left\| \mathbf{w}_{\mathrm{r}} \right\| ^2\sigma _{\mathrm{ap}}^{2}},
\label{eq12}
\end{align}
where ($a$) holds because of $\mathbb{E} \left(\widetilde{\mathbf{C}}_{\mathrm{t}} \widetilde{\mathbf{C}}_{\mathrm{t}}^H \right)={\mathbf{I}}_{L}$ and $\mathbf{XX}^H\approx L\mathbf{WW}^H$.
When the channel $\mathbf{h}_{\mathrm{b}}$ is known to the AP, we could use equal gain combining vector, i.e., $\mathbf{w}_{\mathrm{r}}=\mathbf{h}_{\mathrm{b}}^{H}/\left\| \mathbf{h}_{\mathrm{b}} \right\|$.
\begin{Remark}
Since $\mathbf{X}$ is completely known to the AP, the SINR of the received signal at the AP is related to the sample covariance matrix of $\mathbf{X}$, not just related to the signal $\mathbf{w}_{\mathrm{t}}\mathbf{s}_{\mathrm{t}}$, which is different from \cite{Luo2023}. In other words, communication signal $\mathbf{w}_{\mathrm{u}}\mathbf{s}_{\mathrm{u}}$ and dedicated probing stream $\mathbf{W}_{\mathrm{s}}\mathbf{S}_{\mathrm{s}}$  can also help the tag to facilitate backscatter communication.
\end{Remark}
\begin{Remark}
The data stream $\mathbf{s}_{\mathrm{t}}$ contains a sequence that can activate the passive RF tag for data transmission. Therefore, in order to detect the tag and complete communication transmission, the SINRs at the tag and AP must be greater than a certain threshold to meet their respective sensitivity constraints.
\end{Remark}
\setcounter{equation}{13}
\begin{figure*}[t]
\begin{align}
\gamma _\mathrm{u}&=\frac{\mathbb{E} \left( \left\| \mathbf{h}_{\mathrm{u}}\mathbf{w}_{\mathrm{u}}\mathbf{s}_{\mathrm{u}}^{H} \right\| ^2 \right)}{\mathbb{E} \left( \left\| \mathbf{h}_{\mathrm{u}}\mathbf{w}_{\mathrm{t}}\mathbf{s}_{\mathrm{t}}^{H} \right\| ^2 \right) +\mathbb{E} \left( \left\| \sum_{i=1}^{N_t}{\mathbf{h}_{\mathrm{u}}\mathbf{w}_i\mathbf{s}_i^{H}} \right\| ^2 \right) +\mathbb{E} \left( \left\| h_\mathrm{tu}\left[ \sqrt{\alpha}\mathbf{h}_{\mathrm{f}}\left( \mathbf{w}_{\mathrm{u}}\mathbf{s}_{\mathrm{u}}^{H}+\mathbf{w}_{\mathrm{t}}\mathbf{s}_{\mathrm{t}}^{H}+\mathbf{W}_{\mathrm{s}}\mathbf{S}_{\mathrm{s}} +\sigma _{\mathrm{t}}^{2} \right) \odot \mathbf{c}_{\mathrm{t}} \right] \right\| ^2 \right) +\mathbb{E} \left( \left\| \mathbf{n}_{\mathrm{u}} \right\| ^2 \right)}\nonumber
\\
&=\frac{|\mathbf{h}_{\mathrm{u}}\mathbf{w}_{\mathrm{u}}|^2}{|\mathbf{h}_{\mathrm{u}}\mathbf{w}_{\mathrm{t}}|^2+\sum_{i=1}^{N_t}{|\mathbf{h}_{\mathrm{u}}\mathbf{w}_i|^2}+\alpha |h_\mathrm{tu}|^2\left( |\mathbf{h}_{\mathrm{f}}\mathbf{w}_{\mathrm{u}}|^2+|\mathbf{h}_{\mathrm{f}}\mathbf{w}_{\mathrm{t}}|^2+\sum_{i=1}^{N_t}{|\mathbf{h}_{\mathrm{f}}\mathbf{w}_i|^2}+\sigma _{\mathrm{t}}^{2} \right) +\sigma _{\mathrm{u}}^{2}}.
\label{eq14}
\end{align}
\hrulefill
\end{figure*}
\setcounter{equation}{12}
\subsection{UE Communication Model}
Let $\mathbf{h}_{\mathrm{u}}\in \mathbb{C} ^{1\times N_t}$ and $h_{\mathrm{tu}}\in \mathbb{C}$ represent  the channel between the AP and UE and the channel between the tag and UE, respectively. Then, the received communication signal at the UE $\mathbf{y}_{\mathrm{u}}\in \mathbb{C} ^{1\times L}$ can be expressed as
\begin{align}
\mathbf{y}_{\mathrm{u}}=&\mathbf{h}_{\mathrm{u}}\mathbf{X}+h_{\mathrm{tu}}\mathbf{y}_{\mathrm{b}}+\mathbf{n}_{\mathrm{u}}\nonumber
\\
=&\mathbf{h}_{\mathrm{u}}\mathbf{X}+h_\mathrm{tu}\left( \sqrt{\alpha}\mathbf{h}_\mathrm{u}\mathbf{X}\odot  \mathbf{c}_{\mathrm{t}}+\sqrt{\alpha}\mathbf{n}_{\mathrm{t}}\odot  \mathbf{c}_{\mathrm{t}} \right) +\mathbf{n}_{\mathrm{u}}\nonumber
\\
=&\mathbf{h}_{\mathrm{u}}\mathbf{w}_{\mathrm{u}}\mathbf{s}_{\mathrm{u}}^{H}+\mathbf{h}_{\mathrm{u}}\mathbf{w}_{\mathrm{t}}\mathbf{s}_{\mathrm{t}}^{H}+\sum_{i=1}^{N_t}{\mathbf{h}_{\mathrm{u}}\mathbf{w}_i\mathbf{s}_i^{H}}\nonumber
\\&+h_{\mathrm{tu}}\sqrt{\alpha}\mathbf{h}_{\mathrm{f}}\mathbf{w}_{\mathrm{u}}\mathbf{s}_{\mathrm{u}}^{H}\odot  \mathbf{c}_{\mathrm{t}}+h_{\mathrm{tu}}\sqrt{\alpha}\mathbf{h}_{\mathrm{f}}\mathbf{w}_{\mathrm{t}}\mathbf{s}_{\mathrm{t}}^{H}\odot  \mathbf{c}_{\mathrm{t}}\nonumber
\\&+\sum_{i=1}^{N_t}{h_{\mathrm{tu}}\sqrt{\alpha}\mathbf{h}_{\mathrm{f}}\mathbf{w}_i\mathbf{s}_i^{H}\odot \mathbf{c}_t}+h_{\mathrm{tu}}\sqrt{\alpha}\mathbf{n}_{\mathrm{t}}\odot  \mathbf{c}_{\mathrm{t}}+\mathbf{n}_{\mathrm{u}}.
\label{eq13}
\end{align}
There are three parts in signal $\mathbf{y}_{\mathrm{u}}$, containing the signal $\mathbf{h}_{\mathrm{u}}\mathbf{X}$ from the AP, interference $h_{\mathrm{tu}}\mathbf{y}_{\mathrm{b}}$ from tag, and $\mathbf{n}_{\mathrm{u}}$ the noise vector at the UE, where $\mathbf{n}_{\mathrm{u}}$ is assumed as  $\mathbf{n}_{\mathrm{u}}\sim \mathcal C\mathcal N \left( 0,\sigma_{\mathrm{u}}^{2}\mathbf{I}_L \right)$. The SINR of the signal received by UE is given in (\ref{eq14}). Thus, the communication rate of UE can be expressed as
\setcounter{equation}{14}
\begin{gather}
\begin{aligned}
R={\log _2\left( 1+\gamma _\mathrm{u} \right)}
\end{aligned}
\end{gather}

\subsection{Sensing Model}
\label{2C}
 Tag detection aims to determine whether the tag is present in the environment. The signal received by the AP can be used to perform a signal detection process. We formulate the tag detection problem as a hypothesis testing problem as follows,
\begin{align}
\begin{cases}
	\mathcal{H} _0:\tilde{\mathbf{y}}_{\mathrm{ap}}=\mathbf{w}_{\mathrm{r}}\mathbf{N}_{\mathrm{AP}}\\
	\mathcal{H} _1:\tilde{\mathbf{y}}_{\mathrm{ap}}=[\sqrt{\alpha}\mathbf{w}_{\mathrm{r}}\mathbf{h}_{\mathrm{b}}\mathbf{h}_{\mathrm{f}}(\mathbf{X}+\mathbf{n}_{\mathrm{t}})]\odot \mathbf{c}_\mathrm{t}+\mathbf{w}_{\mathrm{r}}\mathbf{N}_{\mathrm{ap}},
\end{cases}
\label{eq16}
\end{align}
where $\mathcal{H} _0$ means there is no backscatter signal from the tag, $\mathcal{H} _1$ means there is tag echo.
According to Neyman-Pearson criterion \cite{Neyman1992}, the following detector can be obtained
\begin{gather}
\begin{aligned}
\mathrm{Re}\left\{ \tilde{\mathbf{y}}_{\mathrm{ap}} (\sqrt{\alpha}\mathbf{w}_{\mathrm{r}}\mathbf{h}_{\mathrm{b}}\mathbf{h}_{\mathrm{f}}\mathbf{X})^H \right\} 
 \begin{array}{c}
	\overset{\mathcal{H}_1}{\geqslant}\\
	\underset{\mathcal{H}_0}{<}\\
\end{array}\,\, \eta,
 \label{eq17}
\end{aligned}
\end{gather}
where $\eta$ is the detection threshold. The detection probability of the tag denoted as $P_D$ is given by \cite{Tang2022}
\begin{gather}
\begin{aligned}
P_D=\frac{1}{2}\mathrm{erfc}\left\{ \mathrm{erfc}^{-1}\left( 2P_{F} \right) -\sqrt{\gamma _{\mathrm{ap}}} \right\},
\label{eq18}
\end{aligned}
\end{gather}
where $\mathrm{erfc}\left( x \right) =\frac{2}{\sqrt{\pi}}\int_x^{\infty}{e^{-t^2}}dt$ is  the complementary error function. $P_{F}$ is the probability of false alarm. According to \eqref{eq18}, the tag detection probability $P_D$ is a monotonically increasing function of $\gamma _{\mathrm{ap}}$. Therefore, the SINR $\gamma _{\mathrm{ap}}$ of the echo signal can actually represent the ability to detect the tag of the B-ISAC system. In other words, the uplink communication capability of the tag is directly proportional to the ability of the tag detection.

\section{Joint Beamforming Scheme for communication rate Optimization} \label{sec3}
In this section, we focus on the beamforming design for optimizing the communication rate of the UE. we consider the case that the AP has detected the tag and has obtained the information of the channel $\mathbf{h}_{\mathrm{b}}$ and $\mathbf{h}_{\mathrm{f}}$. We consider the problem of maximizing the communication rate of communication UE under the constraints of energy budget, SINRs at the tag and AP. The SINR constraint at the tag ensures that the tag remains activated, while the SINR constraint at the AP ensures the detection probability and uplink communication rate of the tag.
The optimization problem is given by
\begin{subequations}
 \begin{align}
\left( \mathcal{P} _1 \right)~~  & \mathop{\mathrm{maximize}} \limits_{\,\,\mathbf{W}}   &&\log _2\left( 1+\gamma _{\mathrm{u}} \right) \\
&~\mathrm{subject~to}   &&\gamma _{\mathrm{t}}\geqslant \gamma _{\mathrm{tth}}
\label{eq19b}
\\
& &&\gamma _{\mathrm{ap}}\geqslant \gamma _{\mathrm{apth}}
\label{eq19c}
\\
& &&\mathrm{Tr}\left( \mathbf{WW}^H \right) \leqslant P_T,
\end{align}
\end{subequations}
 where $\gamma _{\mathrm{t}}\geqslant \gamma _{{\mathrm{tth}}}$ is to set a threshold $\gamma _{{\mathrm{tth}}}$ for the to ensure the tag be activated,  $\gamma _{\mathrm{ap}}\geqslant \gamma _{{\mathrm{apth}}}$ is to set a threshold $\gamma _{{\mathrm{apth}}}$ to ensure the uplink communication rate and remain the tag be detected. $P_T$ is the total transmit power, $\mathrm{Tr}\left( \mathbf{WW}^H \right) \leqslant P_T$ is a total power budget constraint for joint beamforming matrix.
 
Since only a single communication UE is considered, optimizing the rate is equivalent to optimizing the communication SINR $\gamma _{\mathrm{u}}$. However, according to \eqref{eq14}, $\gamma _{\mathrm{u}}$ has a complicated fractional form. By introducing an auxiliary variables $y$, we can convert the fractional objective function into polynomial form by quadratic transform \cite{Shen2018}. The new objective function $\mathcal{F} \left( \mathbf{W},y \right)$ is given by \eqref{eq20}. Then the optimization problem can be expressed as
\begin{figure*}[ht]
\begin{gather}
\begin{aligned}
\mathcal{F} \left( \mathbf{W},y \right) =2y\mathrm{Re}\left\{ \mathbf{h}_{\mathrm{u}}\mathbf{w}_{\mathrm{u}} \right\} -y^2\left[ |\mathbf{h}_{\mathrm{u}}\mathbf{w}_{\mathrm{t}}|^2+\sum_{i=1}^{N_t}{|\mathbf{h}_{\mathrm{u}}\mathbf{w}_i|^2}+\alpha |h_{\mathrm{tu}}|^2\left( |\mathbf{h}_{\mathrm{f}}\mathbf{w}_{\mathrm{u}}|^2+|\mathbf{h}_{\mathrm{f}}\mathbf{w}_{\mathrm{t}}|^2+\sum_{i=1}^{N_t}{|\mathbf{h}_{\mathrm{f}}\mathbf{w}_i|^2}+\sigma _{\mathrm{t}}^{2} \right) +\sigma _{\mathrm{u}}^{2} \right] .
\label{eq20}
\end{aligned}
\end{gather}
\hrulefill
\end{figure*}
\setcounter{equation}{22}
\begin{figure*}[ht]
\begin{gather}
\begin{aligned}
y^*=\frac{\mathrm{Re}\left\{ \mathbf{h}_{\mathrm{u}}\mathbf{w}_{\mathrm{u}} \right\}}{|\mathbf{h}_{\mathrm{u}}\mathbf{w}_{\mathrm{t}}|^2+\sum_{i=1}^{N_t}{|\mathbf{h}_{\mathrm{u}}\mathbf{w}_i|^2}+\alpha |h_\mathrm{tu}|^2\left( |\mathbf{h}_{\mathrm{f}}\mathbf{w}_{\mathrm{u}}|^2+|\mathbf{h}_{\mathrm{f}}\mathbf{w}_{\mathrm{t}}|^2+\sum_{i=1}^{N_t}{|\mathbf{h}_{\mathrm{f}}\mathbf{w}_i|^2}+\sigma _{\mathrm{t}}^{2} \right) +\sigma _{\mathrm{u}}^{2}}.
\label{eq23}
\end{aligned}
\end{gather}
\hrulefill
\end{figure*}
\setcounter{equation}{20}
\begin{subequations}
 \begin{align}
\left( \mathcal{P} _{1.1} \right)~~  & \mathop{\mathrm{maximize}} \limits_{\,\,\mathbf{W},y}   &&\mathcal{F} \left( \mathbf{W},y \right)
\\
&~\mathrm{subject~to}   &&\gamma _{\mathrm{t}}\geqslant \gamma _{\mathrm{tth}}
\label{eq40b}
\\
& &&\gamma _{\mathrm{ap}}\geqslant \gamma _{\mathrm{apth}}
\label{eq40c}
\\
& &&\mathrm{Tr}\left( \mathbf{WW}^H \right) \leqslant P_T,
\end{align}
\end{subequations}
where $\mathcal{F} \left( \mathbf{W},y \right)$  is a conditionally concave
function with respect to each variable given the other. Therefore, we can develop an alternating optimization method to solve this problem.

\textbf{Update $y$}: Given $\mathbf{W}$, the optimization for the auxiliary variables $y$ is a convex problem without constraints, given as
\begin{gather}
\begin{aligned}
\label{eq41}
\left( \mathcal{P} _{1.1.1} \right)~~  & \mathop{\mathrm{maximize}} \limits_{\,\,y}   &&\mathcal{F} \left( \mathbf{W},y \right).
\end{aligned}
\end{gather}
Its optimal solution can be obtained straightforwardly by setting $\frac{\partial \mathcal{F}}{\partial y}=0$. The optimal $y^*$ is given by \eqref{eq23}.

\textbf{Update $\mathbf{W}$}: Given $y$, the optimization problem for updating $\mathbf{W}$ can be expressed as 
\setcounter{equation}{23}
\begin{subequations}
 \begin{align}
\left( \mathcal{P} _{1.1.2} \right)~~ & \mathop{\mathrm{maximize}} \limits_{\,\,\mathbf{W}=\left[\mathbf{w}_{\mathrm{u}},\mathbf{w}_{\mathrm{t}},\mathbf{w}_{1},\ldots,\mathbf{w}_{N_t} \right] }   &&\mathcal{F} \left( \mathbf{W},y \right)
\\
&~~~~~~~~\mathrm{subject~to}   &&\gamma _{\mathrm{t}}\geqslant \gamma _{\mathrm{tth}}
\label{eq24b}
\\
& &&\gamma _{\mathrm{ap}}\geqslant \gamma _{\mathrm{apth}}
\label{eq24c}
\\
& &&\mathrm{Tr}\left( \mathbf{WW}^H \right) \leqslant P_T.
\end{align}
\end{subequations}
Note that the objective function is concave with respect to $\mathbf{W}$. The main challenge is the non-convex constraints  (\ref{eq24b}) and (\ref{eq24c}). According to \eqref{eq8}, the
(\ref{eq24b}) can be rewritten as
\begin{gather}
\begin{aligned}
\label{eq25}
\frac{1}{\gamma _{\mathrm{tth}}}|\mathbf{h}_{\mathrm{f}}\mathbf{w}_{\mathrm{t}}|^2\geqslant |\mathbf{h}_{\mathrm{f}}\mathbf{w}_{\mathrm{u}}|^2+\sum_{i=1}^{N_t}{|\mathbf{h}_{\mathrm{f}}\mathbf{w}_i|^2}+\sigma _{\mathrm{t}}^{2}.
\end{aligned}
\end{gather}
By taking root square of both side of \eqref{eq25}, the constraint becomes a second-order cone form, which is given by
\begin{gather}
\begin{aligned}
\label{eq26}
\sqrt{\frac{1}{\gamma _{\mathrm{tth}}}}\mathrm{Re}\left\{ \mathbf{h}_{\mathrm{f}}\mathbf{w}_{\mathrm{t}} \right\} \geqslant \left\| \begin{array}{c}
	\mathbf{h}_{\mathrm{f}}\mathbf{w}_{\mathrm{u}}\\
	\mathbf{h}_{\mathrm{f}}\mathbf{w}_1\\
	\mathbf{h}_{\mathrm{f}}\mathbf{w}_2\\
	\vdots\\
	\mathbf{h}_{\mathrm{f}}\mathbf{w}_{N_t}\\
	\sigma _{\mathrm{t}}\\
\end{array} \right\| .
\end{aligned}
\end{gather}
The original form of left hand side of \eqref{eq25} after taking root square is $\sqrt{\frac{1}{\gamma _{\mathrm{tth}}}}| \mathbf{h}_{\mathrm{f}}\mathbf{w}_{\mathrm{t}}|$. While we find that any phase rotation does not affect the absolute value, i.e., $| \mathbf{h}_{\mathrm{f}}\mathbf{w}_{\mathrm{t}}|= | \mathbf{h}_{\mathrm{f}}\mathbf{w}_{\mathrm{t}}e^{j\theta}|$. Without changing the absolute value, we can always find a $\mathbf{w}_{\mathrm{t}}$  to make the $\mathbf{h}_{\mathrm{f}}\mathbf{w}_{\mathrm{t}}$ positive and real. Therefore, the constraint \eqref{eq25} can be rewritten as \eqref{eq26}.

According to \eqref{eq12}, the constraint \eqref{eq24c} can be rewritten as 
 \begin{gather}
\begin{aligned}
\label{eq27}
\frac{\alpha}{\gamma _{\mathrm{apth}}}|\mathbf{w}_{\mathrm{r}}\mathbf{h}_{\mathrm{b}}|^2  
\mathrm{Tr}\left( \mathbf{F}\mathbf{WW}^H \right) \geqslant \left\| \begin{array}{c}
	 \sqrt{\alpha}|\mathbf{w}_{\mathrm{r}}\mathbf{h}_{\mathrm{b}}|\sigma _{\mathrm{t}}\\
	\left\| \mathbf{w}_{\mathrm{r}} \right\|\sigma_{\mathrm{ap}}\\
\end{array} \right\|^2 .
\end{aligned}
\end{gather}
where $\mathbf{F}=\mathbf{h}_\mathrm{f}^{H}\mathbf{h}_\mathrm{f}$.
Now, the main limitation for solving this problem is that constraint \eqref{eq27} is not an affine constraint. In order to address this issue, we adopt successive convex approximation (SCA) based method \cite{Scutari2014} to solve it. Given a $\mathbf{W}^{\ddagger}$, the convex approximation of \eqref{eq27} at $\mathbf{W}^{\ddagger}$ is given by
 \begin{gather}
\begin{aligned}
\label{eq28}
&\frac{\alpha}{\gamma _{\mathrm{apth}}}|\mathbf{w}_{\mathrm{r}}\mathbf{h}_{\mathrm{b}}|^2 \mathrm{Tr}\left( \mathbf{F}\mathbf{W}^{\ddagger}\mathbf{W}^{\ddagger H} \right)\\
+&\frac{\alpha}{\gamma _{\mathrm{apth}}}|\mathbf{w}_{\mathrm{r}}\mathbf{h}_{\mathrm{b}}|^2\mathrm{Tr}\left[ \mathbf{W}^H\mathbf{F}(\mathbf{W}-\mathbf{W}^{\ddagger})+\mathbf{W}^T\mathbf{F}^T(\mathbf{W}-\mathbf{W}^{\ddagger})^* \right]
\\ \geqslant &\left\| \begin{array}{c} \sqrt{\alpha}|\mathbf{w}_{\mathrm{r}}\mathbf{h}_{\mathrm{b}}|\sigma _{\mathrm{t}}\\
	\left\| \mathbf{w}_{\mathrm{r}} \right\|\sigma_{\mathrm{ap}}\\
\end{array} \right\|^2 .
\end{aligned}
\end{gather}

The convex approximation problem of $\left( \mathcal{P} _{1.1.2.1} \right)$ can be written as follows,
\begin{subequations}
 \begin{align}
\left( \mathcal{P} _{1.1.2.1} \right)~~  & \mathop{\mathrm{maximize}} \limits_{\,\,\mathbf{W}}   &&\mathcal{F} \left( \mathbf{W},y \right)
\\
&~\mathrm{subject~to}   &&\eqref{eq26} \nonumber
\\
& &&\eqref{eq28} \nonumber
\\
& &&\mathrm{Tr}\left( \mathbf{WW}^H \right) \leqslant P_T.
\end{align}
\label{eq29}
\end{subequations}
The SCA based algorithm for solving $\left( \mathcal{P}_{1.1.2.1} \right)$ is summarized in Algorithm \ref{alg2}. Based on the above derivations, the optimal joint beamformer can be obtained by updating $y$ and $\mathbf{W}$ iteratively. The alternating joint beamforming design algorithm for communication rate optimization is summarized in Algorithm \ref{alg3}. With appropriate initialization of $y$ and $\mathbf{W}$ in the feasible space, we could iteratively update each variable until convergence.
\renewcommand{\algorithmicrequire}{ \textbf{Input:}} 
\renewcommand{\algorithmicensure}{ \textbf{Output:}} 
\begin{algorithm}[!t]
    \caption{SCA based algorithm for solving $(\mathcal{P}_{1.1.2.1})$ }
    \label{alg2}
    \begin{algorithmic} [1]
    \REQUIRE \, 
    $P_T$, $\sigma_\mathrm{ap}^2$, $\sigma_\mathrm{t}^2$, $\sigma_\mathrm{u}^2$, $\mathbf{h}_\mathrm{u}$, $\mathbf{h}_\mathrm{f}$, $\mathbf{h}_\mathrm{b}$, $N_t$, $N_r$, $\alpha$, $\gamma_\mathrm{uth}$, $\gamma_\mathrm{tth}$, $\gamma_\mathrm{apth}$, 
    $y$.
    \ENSURE Designed beamforming matrix $\mathbf{W}^{\star}$.
    \renewcommand{\algorithmicensure}{ \textbf{Steps:}}
    \ENSURE \, 
   \STATE Initialize: $\mathbf{W}$, $\delta_\mathrm{th}$, ${I}_{\text{max}}$ , $i=1$, 
    $\delta=\infty$.
   
    \STATE \textbf{while} $i \leqslant {I}_{\text{max}}$ and  $\delta \geqslant \delta_\mathrm{th}$  \textbf{do}
    \STATE ~~~~~Let $\mathbf{W}^{\ddagger}=\mathbf{W}$
    \STATE ~~~~~Update $\mathbf{W}$ by solving  \eqref{eq29}.
    \STATE ~~~~~~ $\delta=|\mathrm{Tr}\left[ \mathbf{W}^H\mathbf{F}(\mathbf{W}-\mathbf{W}^{\ddagger})+\mathbf{W}^T\mathbf{F}^T(\mathbf{W}-\mathbf{W}^{\ddagger})^* \right]|$.
    \STATE ~~~~~$i=i+1$.
    \STATE \textbf{end while}
    \STATE Return $\mathbf{W}^{\star}=\mathbf{W}$. 
\end{algorithmic}
\end{algorithm}
\renewcommand{\algorithmicrequire}{ \textbf{Input:}} 
\renewcommand{\algorithmicensure}{ \textbf{Output:}} 
\begin{algorithm}[!t]
    \caption{Alternating algorithm for sloving $(\mathcal{P}_{1})$ }
    \label{alg3}
  \begin{algorithmic} [1]
    \REQUIRE \, 
    $P_T$, $\sigma_\mathrm{ap}^2$, $\sigma_\mathrm{t}^2$, $\sigma_\mathrm{u}^2$, $\mathbf{h}_\mathrm{u}$, $\mathbf{h}_\mathrm{f}$, $\mathbf{h}_\mathrm{b}$, $N_t$, $N_r$, $\alpha$, $\gamma_\mathrm{uth}$, $\gamma_\mathrm{tth}$, $\gamma_\mathrm{apth}$.
    \ENSURE Designed beamforming matrix $\mathbf{W}^{*}$.
    \renewcommand{\algorithmicensure}{ \textbf{Steps:}}
    \ENSURE \, 
   \STATE Initialize: $\mathbf{W}$, $\varepsilon_\mathrm{th}$, ${K}_{\text{max}}$ , $k=1$, 
    $\varepsilon=\infty$, $y_0=0$. 
    \STATE \textbf{while} $k \leqslant {K}_{\text{max}}$ and  $\varepsilon \geqslant \varepsilon_\mathrm{th}$  \textbf{do}
    \STATE ~~~~~Update $y_k$ by \eqref{eq23}. 
    \STATE ~~~~~Update $\mathbf{W}$ by using \textbf{Algorithm \ref{alg2}}.
    \STATE ~~~~ $\varepsilon=y_k-y_{k-1}$.
    \STATE ~~~~~$k=k+1$.
    \STATE \textbf{end while}
    \STATE Return $\mathbf{W}^{{*}}=\mathbf{W}$.
\end{algorithmic}
\end{algorithm}

\begin{figure}[t]
\centering
\subfigure[Convergence of Algorithm \ref{alg2}.]{
\includegraphics[width=0.2276\textwidth]{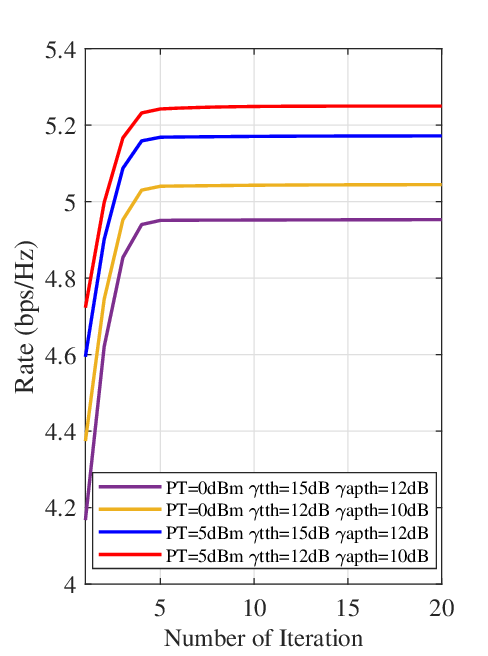}}
\subfigure[Convergence of Algorithm \ref{alg3}.]{
\includegraphics[width=0.2276\textwidth]{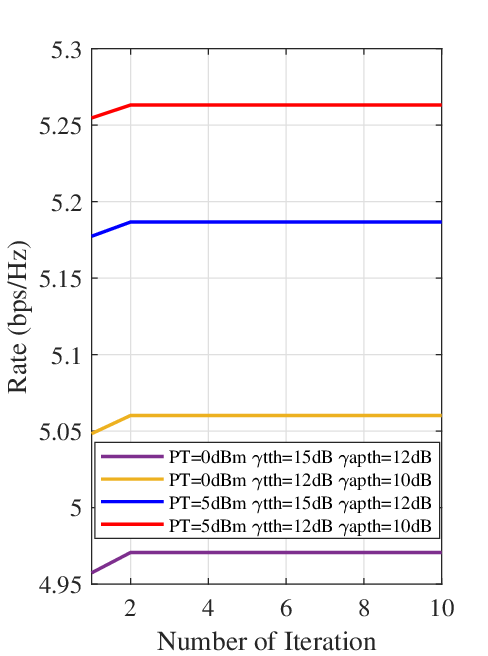}}
\captionsetup{font=small}
\caption{Convergence performance of the proposed Algorithms.}
\label{fig2}
\vspace{-0.4cm}
\end{figure}
\begin{figure}[!t]
\captionsetup{font=small}
\begin{center}
\includegraphics[width=0.4\textwidth, trim=10 1 30 15, clip]{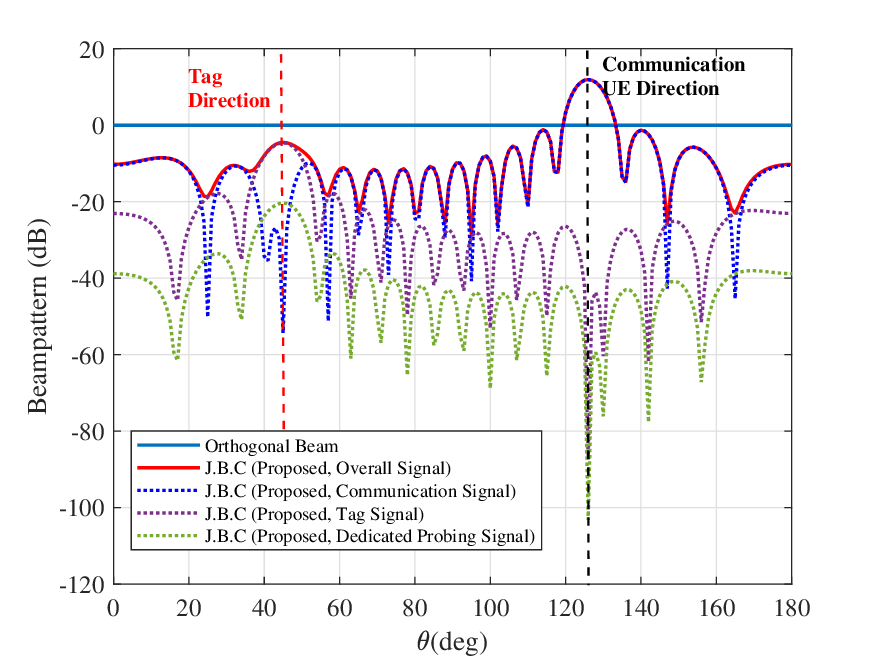}
\end{center}
\vspace{-0.3cm}
\caption{Beampattern of the proposed joint beamforming scheme J.B.C.}
\label{fig3}
\vspace{-0.5cm} 
\end{figure}
\begin{figure}[!t]
\captionsetup{font=small}
\begin{center}
\includegraphics[width=0.4\textwidth, trim=10 1 30 15, clip]{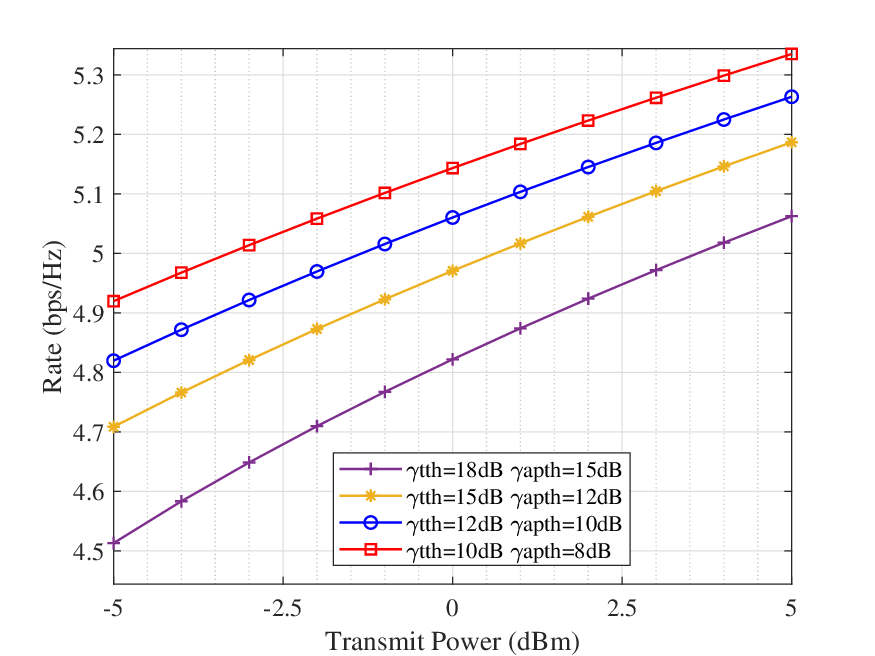}
\end{center}
\vspace{-0.3cm}
\caption{The achievable communication rate of the proposed J.B.C scheme versus the
transmit power}
\label{fig4}
\vspace{-0.7cm} 
\end{figure}

\section{Numerical Results} \label{sec5}
In this section, we present numerical results of the proposed beamforming schemes. We set the transmit and receive array with the same elements number, i.e., $N_t =N_r=16$. We assume that the noise power at the AP, tag, and UE are equal as $\sigma _\mathrm{ap}^{2}=\sigma _\mathrm{t}^{2}=\sigma _\mathrm{u}^{2}=-40$ dBm. We use the line-of-sight (LOS) channel model in the simulation, which means that if the tag is at angle $\theta_k$, the corresponding channel $\mathbf{h}_\mathrm{f}$ is $\alpha_\mathrm{f}\mathbf{a}(\theta_k)$ and the channel $\mathbf{h}_\mathrm{b}$ is $\alpha_\mathrm{b}\mathbf{b}(\theta_k)$. If the UE is at angle $\theta_j$, the corresponding channel $\mathbf{h}_\mathrm{u}$ is $\alpha_\mathrm{u}\mathbf{a}(\theta_j)$. $\mathbf{a}$ and $\mathbf{b}$ are the steering vector of the transmit and receive array respectively. $\alpha_\mathrm{f}$, $\alpha_\mathrm{b}$, and $\alpha_\mathrm{u}$ are channel fading coefficients respectively.

First, we analyze the convergence of the proposed Algorithm \ref{alg2} and Algorithm \ref{alg3}. The convergence performance of the algorithms are presented in Fig.~\ref{fig2}(a) and Fig.~\ref{fig2}(b), respectively. The achievable communication rate versus the number of iterations under different setting is present in the figure. Algorithm \ref{alg2} converges after 10 iterations. In particular, after 5 iterations, the change of rate is already minor. Algorithm \ref{alg3} converges after 5 iterations. In particular, after 2 iterations, the change of rate is already minor. The above simulation results fully demonstrate the quick convergence performance of the proposed algorithms.

Then, we evaluate the beampattern of the proposed joint beamforming scheme for communication rate optimization (J.B.C). We set the transmit power of the AP as $P_T=0$ dBm, and set $h_{\mathrm{tu}}=0.5$. We set the channels to $\mathbf{h}_\mathrm{f}=0.8\mathbf{a}(\frac{\pi}{4})$, $\mathbf{h}_\mathrm{b}=0.8\mathbf{b}(\frac{\pi}{4})$, and  $\mathbf{h}_\mathrm{u}=0.8\mathbf{a}(\frac{7\pi}{10})$. The SINR threshold at the tag is set to $\gamma_\mathrm{tth}=15$dB to ensure the tag is activated. The SINR threshold at the AP is set to $\gamma_\mathrm{tth}=12$dB to make sure the tag can be detected and communicate with the AP. The beampattern of the proposed J.B.C scheme is presented in Fig.~\ref{fig3}. J.B.C (Proposed, Overall Signal), J.B.C (Proposed, Communication Signal), J.B.C (Proposed, Tag Signal), and J.B.C (Proposed, Dedicated Probing Signal) are beampatterns of the overall signal, the communication signal, the tag signal, and the dedicated probing signal produced by the proposed J.B.C scheme (solving problem  $\left( \mathcal{P} _1 \right)$) respectively. Orthogonal Beam is the full orthogonal beamforming scheme. We can observe that the communication beam forms a high gain beam at the communication UE direction and a notch at the tag direction. In order to maximize the UE communication rate, the tag signal and dedicated probing signal produce a high-gain beam in the tag direction and a notch in the communication direction.

The achievable communication rate with respect to the
transmit power under different settings is depicted in Fig.~\ref{fig4}. We can also find that the communication performance of the proposed J.B.C scheme increases with the transmit power. We can also observe that the higher the SINR constraints, the lower the achievable communication rate, which reflects the power competition between the tag and UE.

In this work, we only consider the UE communication performance optimization in B-ISAC systems. The design for more task modes of B-ISAC systems is studied in \cite{Zhaoz2024}. In addition, the design of B-ISAC systems with multiple UEs and multiple RF tags is also a very interesting topic, which we leave as our future work.

\section{Conclusion} \label{sec6}
We proposed an integrated BackCom and ISAC system called B-ISAC system in this work. We provided a theoretical analysis of the communication and sensing performance of the system. A joint beamforming scheme is designed to optimize the UE communication rate under the constraints of energy budget, SINRs at the tag and AP to ensure the tag sensing and communication performance. Moreover, efficient algorithms were developed for solving the complicated optimization problem. Simulation results validate the effectiveness of the proposed algorithms as well as illustrating the performance trade-off between communication and sensing performance.

\section*{Acknowledgment}
The work was supported in part by the National Natural Science Foundation of China under Grant 62388102, and the GuangDong Basic and Applied Basic Research Foundation under Grant 2022A1515010209. The corresponding author is Dr. Yuhan Dong.

\end{document}